\documentclass{aastex}
\begin{document}
\title{An Analytical Approach to the Spin Distribution of
Dark Halos}
\author{Tzihong Chiueh}
\affil{Department of Physics, National Taiwan University,
Taipei, Taiwan}
\email{chiuehth@phys.ntu.edu.tw}
\author{Jounghun Lee}
\affil{Institute of Astronomy and Astrophysics, Academia Sinica,
Taipei, Taiwan}
\email{taiji@asiaa.sinica.edu.tw}
\and
\author{Lihwai Lin}
\affil{Department of Physics, National Taiwan University, Taipei,
Taiwan} \email{d90222005@ms90.ntu.edu.tw}

\begin{abstract}

We derive the distribution of the dimensionless specific angular
momentum of dark matter halos, $P(j)$, in the framework of the
standard tidal torque theory, and explain the characteristic shape
of $P(j)$ commonly observed in N-body simulations. A scalar
quantity (shear scalar), $r$, is introduced for measuring the
effective strength of the tidal torque force acting upon the halo
from the surrounding matter. It is found that the ubiquitous and
broad shape of $P(j)$ can mostly be attributed to the unique
property of the shear scalar $r$, and also that $P(j)$ is
insensitve to the underlying collapse dynamics. Our result
demonstrates that although the shape of $P(j)$ is ubiquitous and
close to log-normal, the distribution is not exactly log-normal
but decays exponentially at the high angular momentum end, and
drops slowly as a power-law with an index $2$ at the low angular
momentum end.
\end{abstract}


\newcommand{\etal}{{\it et al.}}
\newcommand{\beq}{\begin{equation}}
\newcommand{\eeq}{\end{equation}}
\newcommand{\ben}{\begin{eqnarray}}
\newcommand{\een}{\end{eqnarray}}
\newcommand{\lam}{\lambda}
\newcommand{\sig}{\sigma}
\newcommand{\del}{\delta}
\newcommand{\br}{{\bf r}}
\newcommand{\bR}{{\bf R}}
\newcommand{\bq}{{\bf q}}
\newcommand{\bI}{{\bf I}}
\newcommand{\bT}{{\bf T}}
\newcommand{\bL}{{\bf L}}
\newcommand{\tbL}{\tilde{\bf L}}
\newcommand{\tL}{j}
\newcommand{\tI}{\tilde{I}}
\newcommand{\tbI}{\tilde{\bf I}}
\newcommand{\cI}{\bar{I}}
\newcommand{\cbI}{\bar{\bf I}}
\section{INTRODUCTION}

N-body simulations have shown that the dark matter halos of the universe
have the intrinsic angular momentum, the distribution of which has a broad
and ubiquitous shape independent of other halo properties such as
the halo mass scale, the halo formation epoch, and so on
\cite{bar-efs87,hea-pea88,lem-kau99,bul-etal01,che-jin02}. This
characteristic distribution of the halo angular momentum is often
approximated by the log-normal distribution with ubiquitous
mean and variance in the log.
Although it has been proved that this characteristic distribution can be
produced by the current standard model \cite{mal-etal01},
the following important questions have yet to be answered: what the
physical origin of the characteristic distribution is, and whether it
is truly log-normal.

The standard model for the origin and evolution of the halo angular
momentum is the tidal torque theory \cite{dor70,whi84}.
In this model, the angular momentum of a proto-halo, $\bL$, is generated by
the misalignment of the principal axes of the local shear tensor with that
of the inertia tensor and grows with time in the
Lagrangian coordinate:
$L_i(t) = - a^{2}(t)\dot{D}(t)\epsilon_{ijk} T_{jl}I_{lk}$.  Here,
$a(t)$ is the scaling factor, $\dot{D}(t)$ the growing rate of
density
perturbations, and $I_{lk}(=\int \rho_{0}q_l q_k d^{3}\bq$ with $\rho_0$
being the mean density) is the
inertia tensor of a proto-halo site in the Lagrangian $\bq$ space,
representing the geometrical shape of a proto-halo.  In addition,
$T_{jl}(=\partial_j\partial_l\phi$ with $\phi$ being the gravitational
potential) is the gravitational shear
tensor that quantifies the tidal torque from the surrounding matter.

Strictly speaking, the tidal torque theory is a linear theory,
and should break down in the non-linear regime since
it does not account for the complicated nonlinear process of halo
formation after recollapse.  To one's surprise, however,
it has been demonstrated by N-body simulations that the linear
tidal torque theory works well even in nonlinear regime
\cite{fre-etal88,hea-pea88,cat-the96,sug-etal00,mal-etal01}.
Therefore, we will adopt the tidal torque model, and
derive the distribution of the halo angular momentum.

\section{DERIVATION}

\subsection{SEPARATION OF THE VARIABLES}

For the distribution of halo angular momentum, it is more
convenient to use a dimensionless quantity than to use $L$ itself
since $L$ scales as $M^{5/3}$:
$L \sim I \sim \int q^{2} d^{3}\bq \sim R^{5} \sim M^{5/3}$, with
$R$ and $M$ being the typical size and mass of a proto-halo respectively.
We thus define a time-independent specific angular momentum, $\tL$,
as $\tL_i\equiv\epsilon_{ijk}T_{jl}\cI_{lk}$ with $\cbI \equiv \bI
M^{-5/3})$, since
the $\log({\tL})$-distribution is evaluated at the same epoch and has a
same shape
as the $\log(L)$-distribution, except for a constant horizontal shift.

Rotating the coordinates into the local shear principal axis frame,
we find that the magnitude of the specific angular momentum is given as
\beq
\tL = \{ \cI_{23}^{2}(\lam_2 - \lam_3 )^{2} +
\cI_{31}^{2}(\lam_1 - \lam_3 )^{2} + \cI_{12}^{2}(\lam_1 - \lam_2 )^{2}
\}^{1/2},
\label{eqn:tidal}
\eeq
where $\lam_1,\lam_2,\lam_3$ are the three eigenvalues of the local shear
tensor, and $\cI_{23},\cI_{31},\cI_{12}$ are the three off-diagonal
components of the dimensionless inertia tensor, $\bar{\bf I}$. 
It is worth noting
that only the off-diagonal components of the inertia tensor in the shear
principal axes are involved in the generation of the halo angular momentum.
The diagonal components of the inertia tensor are involved in the
gravitational collapse.  Interestingly enough, however, it has been recently
found by N-body simulations that the shear and inertia tensors are
strongly, but not perfectly, aligned
\cite{lee-pen00,por-etal01}. In other words, in the shear principal axes,
the magnitudes of the inertia off-diagonal components are small relative
to the diagonal parts. This numerical evidence for the
small off-diagonal components of the inertia tensor prompts us
to regard the misalignment to be statistical fluctuation and
$\cI_{23},\cI_{31},\cI_{12}$ as independent random
variables. Now that
the three eigevalues of the shear tensor are also random variables, and
equation (\ref{eqn:tidal}) describes how the positive random variable,
$\tL$, depends on the random variabes, $\bar{\bf I}$ and $\lam_i$'s.

To derive the distribution of $\tL$, we first separate the
variables as follows.  We view the generation of the halo angular
momentum as a $3\times 3$ diagonal tidal tensor
$R_{ij} \equiv Diag[r_{1},r_{2},r_{3}]$ acting on an 
inertia-moment vector
$\tbI$, i.e., $\tL_i = \sqrt{3}R_{ij}\tI_j$, in the frame of local
shear-principal-axis, where the components of $R_{ij}$ and
$\tbI$ are given as $r_1 = (\lam_2 - \lam_3)/\sqrt{3}$, $r_2 =
(\lam_3 - \lam_1)/\sqrt{3}$ and $r_3 = (\lam_1 -
\lam_2)/\sqrt{3}$, and $\tI_1 \equiv \cI_{23}$, $\tI_2 \equiv
\cI_{31}$ and $\tI_2 \equiv \cI_{12}$. As the three components of
$\tbI$ are independent random variables, we may further express
them in spherical-polar coordinates as $\tI_1 =
\tI\sin\theta\cos\phi$, $\tI_2 = \tI\cos\theta$, $\tI_3 =
\tI\sin\theta\sin\phi$ with $\tI\equiv (\tI_1^{2} + \tI_2^{2} +
\tI_3^{2})^{1/2}$, $\theta\equiv \cos^{-1}(\tI_2/\tI)$,
$\phi\equiv \tan^{-1}(\tI_3/\tI_1)$.  On the other hand the
diagonal components of $R_{ij}$ are not all free but constrained
by the condition of $r_1 + r_2 + r_3 = 0$. One may conveniently
regard the diagonal components $(r_1, r_2, r_3)$ also as a vector
$\br$; this vector lies in a plane perpendicular to the $(1,1,1)$
symmetric axis in the ${\bf\lambda}$ space, or
$\br={\bf\lambda}\times (1,1,1)/\sqrt{3}$. Therefore, $\br$
depends only on one polar angle, say $\psi$. To find the
polar-coordinate expression for $\br$, we perform a coordinate
transformation from $\lam_1,\lam_2,\lam_3$ into $x,y,\delta$, with
$x\equiv(\lam_1 - 2\lam_2 + \lam_3)/\sqrt{6}$, $y\equiv (\lam_1 -
\lam_3)/\sqrt{2}$, $\delta\equiv\lam_1 + \lam_2 + \lam_3$. Under
this transformation, $\br$ has only two components: $\br =
(x,y,0)$. So, we can express $x = r\cos\psi$, $y=r\sin\psi$.  Now,
rotating the axes back into the shear principal axes, we have a
polar-angle expression for $\br$ in the shear principal axis
frame: $r_1= -\frac{1}{\sqrt{2}}r\cos\psi +
\frac{1}{\sqrt{6}}r\sin\psi$, $r_2 = -
\frac{\sqrt{2}}{\sqrt{3}}r\sin\psi$, $r_3 =
\frac{1}{\sqrt{2}}r\cos\psi + \frac{1}{\sqrt{6}}r\sin\psi$.

Using these angular variables for $\tbI$ and $\br$, we find the
following expression for $\tL$ with separated variables: \beq \tL
= \tI r Y(\psi,\theta,\phi), \label{eqn:tj} \eeq where \beq
Y(\psi,\theta,\phi)\equiv
\bigg\{\sin^{2}\theta(\frac{1}{2}\cos^{2}\psi +
\frac{1}{6}\sin^{2}\psi -\frac{1}{2\sqrt{3}}\cos 2\phi\sin 2\psi)
+\frac{2}{3}\cos^{2}\theta\sin^{2}\psi\bigg\}^{1/2}. \eeq The
off-diagonal components of the inertia tensor $\tI_1$, $\tI_2$ and
$\tI_3$ are further assumed to be Gaussian variables
\cite{cat-the96}. So, the distribution of $\tI = \sqrt{\tI_{1}^{2}
+ \tI_{2}^{2} + \tI_{3}^{2}}$ is a weighted Gaussian given as
$P_{\tI}(\tI)d\tI =\sqrt{2/\pi} \sig_{\tI}^{-3}
\tI^2\exp\left(-\tI^{2}/2\sig_{\tI}^{2}\right)d\tI$, where
$\sig_{\tI}$ is the standard deviation of each off-diagonal
component.  While the solid angles, $\phi$ and $\theta$ are
uniformly distributed on the sphere.

Though the shear scalar $r$ was first pointed out to play an
important role for non-spherical halo formation and angular
momentum generation by Chiueh \& Lee (2001), Sheth \& Tormen
(2002) recognized that $r$ can in fact be conveniently expressed
as $[\frac{2}{15}\sum_{i=1}^{5} y_{i}]^{1/2}$ where
$y_1,\cdots,y_5$ are mutually independent Gaussian variables,
having the same standard deviations equal to that of the density
field, $\sig_{\del}$. Therefore, the distribution of $r$ is also a
weighted Gaussian: $P_r(r) = \sqrt{5/\pi}(25/12\sig_{\del}^5)
r^4\exp\left(-5r^2/4\sig^{2}_{\del}\right)$. Sheth \& Tormen
(2002) commented that the broadness of $P(\tL)$ may be attributed
to the shape of $P_r(r)$.  However, $P(\tL)$ is contributed by the
conditional distribution of $r$ only for those regions out of
which dark halos condense, i.e., $P_r(r|{\rm halo})$, instead of
the unconditional $P_r(r)$ given above. In the Press-Schechter
model of halo formation \cite{pre-sch74} where the collapse
threshold depends only on $\del$ and not on $r$, $P_r(r|{\rm
halo})$ indeed equals $P(r)$. However, the actual halo collapse is
generically non-spherical, which leads to higher collapse
thresholds depending both on $\del$ and $r$
\cite{she-etal01,chi-lee01}. This fact differentiates the actual
halo $r$-distribution per mass bin, $P_r(r|{\rm halo})$, from the
unconditional $P_r(r)$.  

To find $P_r(r|{\rm halo})$, we examine the non-spherical collapse
using the collapse condition given by Chiueh \& Lee (2001):
$\frac{\del}{\del_c} = \left( 1 + \frac{r^4}{\alpha}\right)^{\beta}$.
Chiueh \& Lee (2001) originally
suggested $\alpha = \beta = 0.15$; later Lin \etal\ (2001) refined
the values as $\alpha = 0.26$ and $\beta=0.16$, which are adopted here.
Since the collapse condition depends on $r$,
$P_r(r|{\rm halo})$ should no longer equal $P_r(r)$.
We calculated $P_r(r|{\rm halo})$ numerically using the random-walk
method described in Lin et al. (2001).  Two non-trivial results are
found.  First, 
$P_r(r|{\rm halo})$ is independent of mass up to the numerical accuracy,
and second, $P_r(r)$ turns out to still be a good
approximation to $P_r(r|{\rm halo})$. Figure 1. compares the numerical
$P_r(r|{\rm halo})$ near the characteristic mass $M_*$ with $P_r(r)$.
The good argreement  suggests that
the distribution of $\tL$ is not sensitive to the detailed dynamics of
halo formation.  This
feature is the origin of the scale-independent {\it shape} of
the halo angular momentum distribution to be shown below.

\subsection{The $\tL$-Distribution}

Now, replace $P_r(r|{\rm halo})$ by the simpler $P_r(r)$.
By equation (\ref{eqn:tj}), the angular distribution $P(\tL)$ can be
derived as follows:
\beq
P(\tL)=\int d\lam_1d\lam_2d\lam_3 \int d\tI\sin\theta d\theta d\phi P_\lam(\lam_1,\lam_2,\lam_3)P_{\tI}
\delta_D(\tL-\tI r Y(\theta,\phi,\psi)),
\label{eqn:tj2}
\eeq
where
\beq
P_\lam(\lam_1,\lam_2,\lam_3) = \frac{3375}{8\sqrt{5}\pi\sig_{\del}^{6}}
\exp\left(-\frac{\del^{2}}{2\sig_{\del}^{2}} -
\frac{5r^2}{4\sig_{\del}^2}\right)
|(\lam_1-\lam_2)(\lam_2-\lam_3)(\lam_1-\lam_3)|,
\label{eqn:joint}
\eeq
and
the Dirac $\delta_D$-function constrains $\tL$ to other variables through
eq.(2).
Note that the last factor in eq.(5) is no more than
$|r_1 r_2 r_3|(\equiv J(r,\psi))$, arising from the Jacobian of the
angular distribution of the shear principal
axes\cite{dor70}.  Thus, $J(r,\psi)=r^3 J(\psi)$, with
$J(\psi)\equiv |(3\cos^{2}\psi-\sin^{2}\psi)\sin\psi|$.  The angular
dependence of
$J(\psi)$ results in $P_\lam$ to contain three null lines,
separated by $60$ degrees, on
the plane where the ${\bf r}$ vector lies.
The null lines correspond to the degeneracies, $\lam_1=\lam_2$,
$\lam_2=\lam_3$ and $\lam_3=\lam_1$.
This feature renders the random angle $\psi$ to distribute
non-uniformly with three-fold symmetry.

Upon changing the variables from $\lam_1,\lam_2,\lam_3$ to
$\delta, r,\psi$, the $\delta$, $\tI$ and $r$ integrals can all be
evaluated analytically, and eq.(4) now becomes \beq P(\tL) =
w^{-1}\tL^3\int \frac{d^{3}\Omega}{Y^4}
K_{1}(\sqrt{\frac{5}{2}}\frac{\tL}{bY}) J(\psi) \label{eqn:fin}
\eeq where $d^3\Omega\equiv \sin\theta d\theta d\phi d\psi$, $K_1$
is the 1st-order modified Bessel function, $b\equiv
\sqrt{\sig_\delta\sig_{\tI}}$, $w$ the normalization factor and
$Y$ is given in eq.(3). As $\log(\tL/b)=\log(\tL)-\log(b)$, the
shape of the $\log(\tL)$-distribution is not affected by the value
of the unknown $b$ although the peak position is. The angular
integration of eq.(6) can only be evaluated numerically and the
solid line of Figure 2 presents the resulting $\tL P(\tL)$. It
shows that though the main body of $P(\tL)$ resembles the
log-normal distribution (dashed line), the distribution is not
exactly log-normal. One may check from eq.(6) by asymptotic
expansion that $P(\tL)$ drops off exponentially at the high
angular momentum limit (cf. Maller \etal\ 2001) and behaves as a
power law ($\sim \tL^{2}$) in the lower angular momentum limit.
The best-fit log-normal to our distribution has a log-width
$\sigma_{\log\tL}=0.63$ (dashed line), which is in fair agreement
with the log-width, $\sigma_{\log\tL}=0.5$, obtained from N-body
simulations reported by Bullock et al. \cite{bul-etal01}.  (Their
definition of $\lambda'$ is equivalent to our $\tL$ modulo a
constant factor.)

For comparison, we have also simulated non-spherical halo collapse
using the random-walk algorithm. When the collapse condition given
in the previous section is satisfied, the shear tensor is
diagonalized (Chiueh and Lee, 2001) and the three principal
eigenvalues $\lam's$ are used to evaluate the halo shear vector
${\bf r}$. The angular momentum of the dark halo is then computed
according to eq.(1).  Figure 2 also plots the resulting $\tL
P(\tL)$ (solid squares). To ensure the accuracy of the our method,
we have also repeated the same procedure for the Press-Schechter
model and computed its $\tL P(\tL)$ (solid triangles), to which
the solid line should be identical. As one can see, all three
results agree with each other, revealing that the angular momentum
distribution is insensitive to the detailed dynamics of halo
formation.

We would like to stress that the unique shape and $\log(\tL)$ width of
the angular-momentum distribution
are the direct outcome of our model.  It involves no fitting parameter.

\section{DISCUSSIONS AND CONCLUSIONS}

To explain the observed characteristic shape for the
angular momentum distribution of
dark halos in N-body simulations, we have adopted the standard tidal
torque theory to study the proto-halo angular momentum $\tL_i$,
which can succinctly be represented by
a random shear $R_{ij}$
acting upon a random vector $\tI_i$.
The diagonalized $R_{ij}$, whose three components are the
mutual differences of the three eigenvalues of the local shear tensor,
quantifies the non-sphericality of the local gravitational potential,
and the vector $\tbI$ characterizes
the misalignment between the matter distribution and local gravitational
potential.
With this picture, we have derived a new expression for the halo
angular momentum, which separates the variables into three parts:
$r,\tI, Y(\psi,\theta,\phi)$.  With
$P_r(r|{\rm halo})$ being shown to
be approximately a weighted Gaussian and assuming
$P_{\tI}(\tI)$ to also be a weighted Gaussian,
we have derived
the angular-momentum distribution, i.e., $P(\tL)$, to be close to
log-normal, with a mass-independent width in fair agreement with that
determined
from simulations.  This "quasi"-log-normal distribution is simply a
consequence of the nonlinear coupling of five random variables,
$r$, $\tI$, $\theta$, $\phi$ and $\psi$.

Though the distribution of $\log(\tL)$ has a mass-independent
shape, the mean of $\log(\tL)$ is predicted to depend on $b$
$(\equiv\sig_{\tI}\sig_{\del}(M))$. As our framework does not
permit the mass dependence of $\sig_{\tI}$ to be determined, this
present work can only account for the ubiquitous {\it shape}, but
not the peak location, of the angular momentum distribution.
Nonetheless, simulations have empirically found that even the mean
value of $\log\tL$ is also mass
independent\cite{bar-efs87,hea-pea88,lem-kau99,bul-etal01,che-jin02}.
Incorporating this finding, we thus demand that
\begin{equation}
\sigma_{\tI}(M)\propto\sigma_{\del}^{-1}(M).
\end{equation}
Such mass dependence of $\sig_{\tI}$ is turned to be
the prediction
of the present work.  That is, the misalignment (per mass) ${\tI}$
has a wider distribution for larger halos than for
smaller halos at the same epoch, since $\sig_{\del}$ is greater
for smaller halos than for larger halos.
This prediction does make qualitative sense,
in that at a given epoch large halos have undergone vigorous
growth with mergers of
sub-halos; it thus makes the
interior matter distribution of large halos less correlated with the
surrounding matter from which the gravitational forces are exerted.
By contrast, small halos have grown mildly
and hence tend to be in a
quiescent state, thus keeping themselves well correlated with the
surrounding matter.
At any rate, this prediction
as well as the Gaussianity assumption of $\tI$ are
both testable by cosmological simulations.

\acknowledgments

This work has been supported by the Taida-ASIAA CosPA Project.
T.C. acknowledges the partial support from the National Science
Council of Taiwan under the grant: NSC90-2112-M-002-026.

\clearpage
\begin{figure}
\plotone{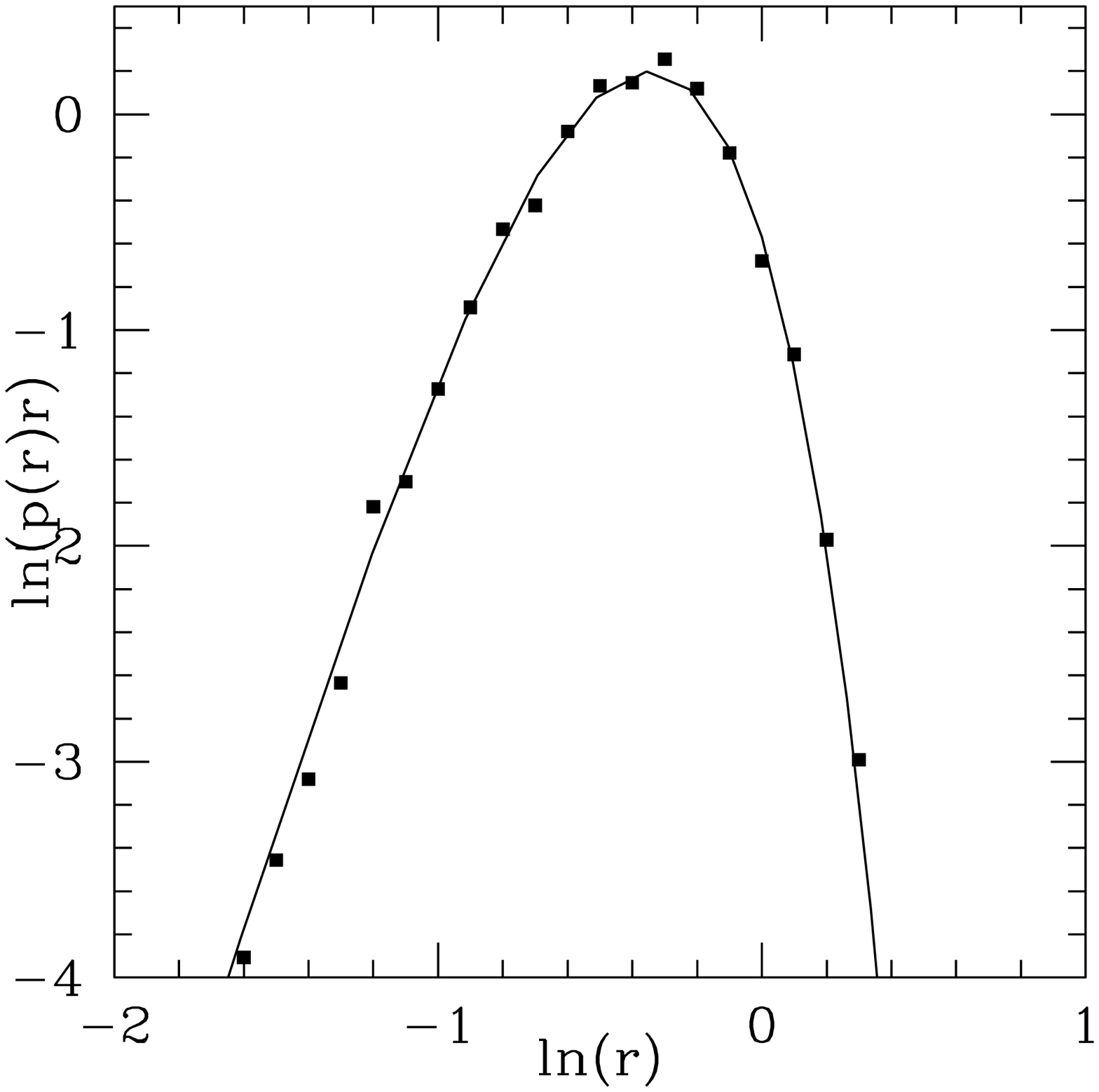}
\caption{The distribution of the halo shear scalar $r$
in the logarithmic
scale. The solid squares are $P_r(r|{\rm halo})$ near $M_*$ obtained
from the Monte-Carlo simulation using the
non-spherical collapse condition.    
The solid curve is $P_r(r)$ given by the
analytic formula with the Press-Schechter collapse condition.}
\end{figure}

\clearpage
\begin{figure}
\plotone{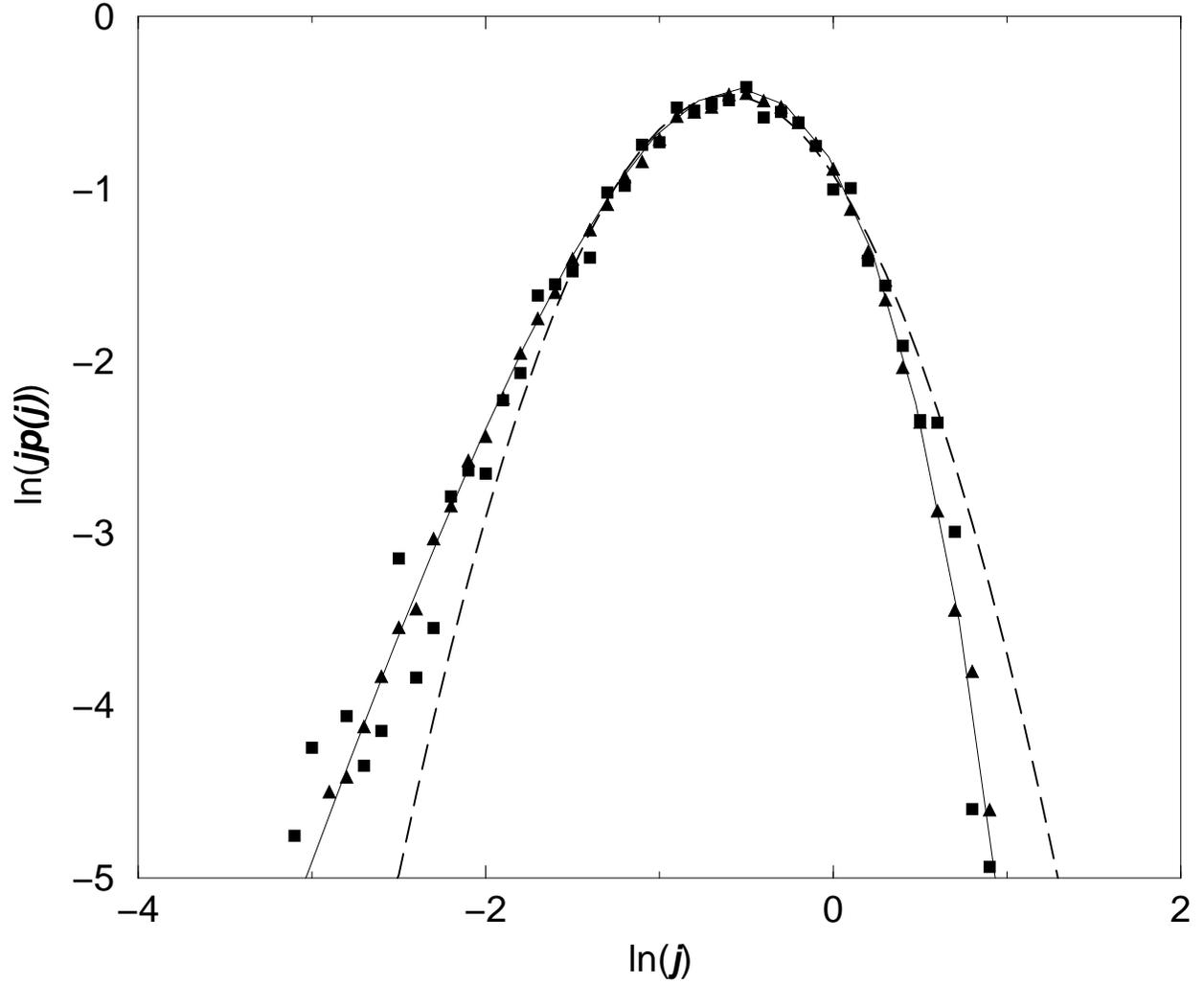} \caption{The distribution of the logarithmic specific
angular momentum of dark halos, $\tL P(\tL)$. The solid curve is
our analytic result given by eq.(6); the squares are Monte Carlo
result for the
non-spherical
dynamical model, and the triangles are Monte Carlo result for the
Press-Schechter model.  The dashed line is the best log-normal fit
to $\tL P(\tL)$.}
\end{figure}
\end{document}